\documentclass[twocolumn,showpacs,preprintnumbers,amsmath,amssymb]{revtex4} 
                                                                                
\usepackage{graphicx}
\usepackage{dcolumn}
\usepackage{bm}
\begin{document}
                                                                                
\preprint{CUPhys/16/2006}
                                                                                
\title{ Spin-spin Correlation in Some Excited States of Transverse Ising
Model }
                                                                                
\author{Anjan Kumar Chandra }
\author{Subinay Dasgupta}%
\affiliation{%
Department of Physics, University of Calcutta,92 Acharya Prafulla Chandra Road, Calcutta 700009, India.\\
}%
                                                                                
\date{\today}
                                                                                
\begin{abstract}

We consider the transverse Ising model in one dimension with nearest-neighbour 
interaction and calculate exactly the longitudinal spin-spin correlation for 
a class of excited states. These states are known to play an important role 
in the perturbative treatment
of one-dimensional transverse Ising model with frustrated second-neighbour
interaction. To calculate the correlation, we follow the earlier procedure of
Wu, use Szego's theorem and also use Fisher-Hartwig conjecture. The result is 
that the correlation decays algebraically with distance ($n$) as $1/\surd n$ 
and is oscillatory or non-oscillatory depending on the magnitude of the 
transverse field. 

\end{abstract}
                                                                                
                                                                                
\def\be{\begin{equation}}
\def\ee{\end{equation}}
\maketitle

\section{Introduction}
Transverse  Axial Next-Nearest Neighbour Ising (TANNNI) model has attracted 
much attention recently as a simple quantum Ising model with frustration.
In one dimension, it is defined by the Hamiltonian \cite{bkc_book,rieger,amit}
\be {\mathcal H}_A = - J \sum_{j=1}^N (s^z_j s^z_{j+1} - 
\kappa s^z_j s^z_{j+2}) - \Gamma \sum_{j=1}^N s^x_j \ee
where $s^z_j$ and $s^x_j$ are the Pauli matrices $\sigma_z$ and $\sigma_x$
at the site $j$, $J$ is the strength of the nearest neighbour 
ferromagnetic interaction,
$\kappa$ is the frustration parameter and $\Gamma$ is the transverse field.
The phase diagram of the ground state of this model in the $\kappa$-$\Gamma$ 
plane shows 
interesting features and is a matter of much controversy and investigation.
Recently, we have shown \cite{cd1} that the ground state of the Hamiltonian 
${\mathcal H}_A$ is related to a class of excited states of a much simpler
Hamiltonian obtained by removing the frustration (i.e. setting $\kappa=0$),
\be
{\mathcal H} =  - J \sum_{j=1}^N s^z_j s^z_{j+1} - \Gamma \sum_{j=1}^N s^x_j.
\ee
This Hamiltonian defines what is called transverse Ising model (in one 
dimension) and is the first quantum Ising model to be solved exactly 
(\cite{LSM,pf}, see also \cite{mattis,bkc_book}). 

All the eigenvalues and eigenfunctions of the Hamiltonian ${\mathcal H}$ 
for transverse Ising model can be derived exactly 
\cite{mattis,bkc_book,LSM,pf}. The $2^N$ number of energy eigenvalues 
are
\be E= 2 \Gamma \sum_k \xi_k \Lambda_k \ee
where $\Lambda_k=\surd(\lambda^2 + 2\lambda \cos k + 1)$, $\lambda=J/\Gamma$,
$\xi_k$  may be 0, $\pm$1 and $k$ runs over $N/2$ equispaced values in the
interval 0 to $\pi$. The ground state is obviously
\[ E_0= - 2 \Gamma \sum_k \Lambda_k \]
corresponding to $\xi_k=-1$ for all $k$ and the excited states correspond to 
the other distributions of $\xi_k$ values. The 
spin-spin correlation function in the longitudinal direction 
\be C^z(n) \equiv < s^z_j s^z_{j+n} > - <s^z_j>^2 \ee
has also been calculated exactly \cite{LSM,pf} for the Hamiltonian
${\mathcal H}$ in the ground state; it decays exponentially with 
distance at $\lambda \ne 1$ and algebraically (with exponent = 1/4) at 
$\lambda = 1$ (the critical point). As for the excited states, the 
correlation was first studied by McCoy \cite{MC}.
The specific quantity calculated was the correlation at a temperature $T > 0$
for the XY model (which in turn can be mapped onto the transverse Ising model).
The correlation was found to decay exponentially with distance except when
$T=0$ and $\lambda = 1$. 

The basic motivation of the paper is that (as mentioned above) the ground 
state of the Transverse  Axial Next-Nearest Neighbour Ising
Hamiltonian ${\mathcal H}_A$ in the region of 
interest (namely, $\kappa \sim 0.5$, $\Gamma \sim 0$, see ref. [4]) 
is related to a particular set of excited states of the transverse 
Ising Hamiltonian ${\mathcal H}$. This states have 
\be \xi_k = \left\{ \begin{array}{rl}
        -1 & \mbox{for $k=0$ to $\phi$ (unexcited)}\\
        1 & \mbox{for $k=\phi$ to $\pi$ (excited)}
        \end{array}
\right. \ee
It is thus important to study these
states for transverse Ising model, especially, to calculate in these states, 
the longitudinal spin-spin correlation
$C^z(n)$ for large $n$. In this paper we provide exact calculation for this
quantity. Our result is that for these states, the correlation 
decays algebraically with exponent 1/2 for all values of $\lambda \ne 1$, 
provided $\phi$ is neither 0 or $\pi$ (some levels are excited, but not all). 
This communication does not cover the case of $\lambda = 1$.

Our results do not contradict the results obtained by McCoy, since at 
$T > 0$, not only the excited states described by Eq. (5) but all possible 
excited states
contribute to the correlation. Thus, any state with energy $E$ would provide
a contribution by an amount proportional to $\exp(-E/k_BT)$ ($k_B$ is the
Boltzmann constant). If the states
with exponentially decaying correlation be dominant, the resultant correlation
will not be algebraic.

Instead of calculating $C^z(n)$ directly, we shall rather calculate the 
quantity
\be \sigma^z(n) \equiv < s^z_j s^z_{j+n} > \ee
We shall see that for the excited states discussed in this paper (see Eq. (5)), 
\[ \lim_{n \rightarrow \infty} \sigma^z(n) = 0\]
so that $<s^z_j>$ is also zero and $C^z(n) = \sigma^z(n).$ 
To evaluate the quantity $\sigma^z(n)$ for large $n$, we note that this 
quantity can be expressed as an $n\times n$ Toeplitz determinant \cite{LSM,pf}
\be
{\det}_n(\lambda) \equiv \left | \begin {array}{ccccc}
  G_0 & G_{-1} & G_{-2} & \cdots & G_{-n+1} \\
  G_1 & G_0 & G_{-1} & \cdots & G_{-n+2} \\
  G_2 & G_1 & G_0 & \cdots & G_{-n+3} \\
  \cdots \cdots & & & & \\
  G_{n-1} & G_{n-2} & G_{n-3} & \cdots & G_0 \\
\end{array}      \right |
\ee
where the elements are given by
\be 
G_m = \frac{1}{\pi} \left[ \int_{k=0}^{\phi} - \int_{k=\phi}^{\pi} 
\right] \frac{dk}{\Lambda_k}[\cos (km-k) + \lambda \cos(km) ] 
\ee
In fact, Pfeuty \cite{pf} has given the expression for $G_m$ in the ground
state and there this integral is from 0 to $\pi$. That expression has
been modified here for the excited state following Lieb, Schultz and 
Mattis \cite{LSM1}.
The standard prescription for studying the limiting ($n \rightarrow \infty$) 
behaviour of a Toeplitz determinant is to apply Szego's Theorem in 
Fisher-Hartwig form \cite{MW,FH,bott,ehr}. We shall also apply the same 
technique on the determinant given above.

In Secs. II and III we treat the cases $\lambda < 1$ and $\lambda > 1$
respectively and in Sec. IV we present some discussions.

\section{Spin Correlations for $\lambda < 1$}

We start this section with some analysis that is true both for $\lambda < 1$ 
and $\lambda > 1$. 
The first step is to observe that the expression (8) for the elements $G_m$
of the determinant can be written as,
\be G_m = \frac{1}{2\pi} \int_0^{2\pi} dk \; \exp (-imk) \; c(k) \ee
with the function $c(k)$ (called the generating function) given by,
\be c(k) = \left\{ \begin{array}{rl}
         w & \mbox{for $0 < k < \phi$}\\
         -w & \mbox{for $\phi < k < 2\pi - \phi$}\\
         w & \mbox{for $2\pi - \phi < k < 2\pi$}\\
         \end{array}
 \right. \ee
where
\be w = \sqrt{\frac{\lambda +e^{ik}}{\lambda +e^{-ik}}}  \ee
It should be mentioned that this generating function is similar to that for
the classical two-dimensional Ising model \cite{MW1} apart from the finite
jumps at $k =\phi$ and $2\pi-\phi$.

Till now, whenever we have come across a square root (Eqs. (3),(8),(11))
we have tacitly adopted
the convention (like usual computer language compilers) that we shall always
take the numerically positive root. However, since later (see Eqs.(16), (17) 
below) we shall need to integrate the logarithm of the generating function 
from 0 to $2\pi$, 
let us choose, following the standard literature \cite{MW5}, the branch of 
square root for which this function is positive in the limit 
$k \rightarrow \pi$. Thus, we write the surd in Eq. (11) as
\be w = \frac{\lambda + \cos k + i \sin k}{\surd(1+\lambda^2+2\lambda\cos k)} 
\tag{11a}\ee
and observe that we should take the negative root for $\lambda > 1$ and
positive root for $\lambda < 1$. Let us call the generating function 
evaluated with this new convention as $\overline{c}(k)$ which is related to
the $c(k)$ of Eq.(10) by
\be \overline{c}(k) = \left\{ \begin{array}{rl}
                     - c(k) & \mbox{for $\lambda > 1$}\\
                     c(k) & \mbox{for $\lambda < 1$}
                    \end{array}
 \right. \ee
This $\overline{c}(k)$ when plugged in Eq. (9) leads to $\overline{G}_m$ and
the Toeplitz determinant ${\overline{\det}}_n(\lambda)$ constructed from 
$\overline{G}_m$ can be 
evaluated by applying Szego's theorem on $\overline{c}(k)$. Obviously, this 
determinant is related to the ${\det}_n(\lambda)$ of Eq. (7) by
\be {\overline{\det}}_n(\lambda) = \left\{ \begin{array}{rl}
                     (-1)^n {\det}_n(\lambda) & \mbox{for $\lambda > 1$}\\
                     {\det}_n(\lambda) & \mbox{for $\lambda < 1$}
                    \end{array}
 \right. \tag{12a}\ee

The ``index'' of the generating function, defined as
\[{\rm Ind} \; \overline{c}(k) = \frac{1}{2\pi i} [\log \overline{c}(2\pi) 
- \log \overline{c}(0)] \]
is 0 for $\lambda < 1$ but +1 for $\lambda > 1$. Since Szego's theorem can be
directly applied only when the index is zero, our job is somewhat simpler 
for $\lambda < 1$, compared to the same for $\lambda > 1$. 

We shall be confined to the case of $\lambda < 1$ henceforth. 
The generating function for this case is,
\be \overline{c}(k) = \left\{ \begin{array}{rl}
                     w & \mbox{for $0 < k < \phi$}\\
                     -w & \mbox{for $\phi < k < 2\pi - \phi$}\\
                     w & \mbox{for $2\pi - \phi < k < 2\pi$}.\\
         \end{array}
 \right. \ee

\begin{figure}
\noindent \includegraphics[width= 6cm, angle = 270]{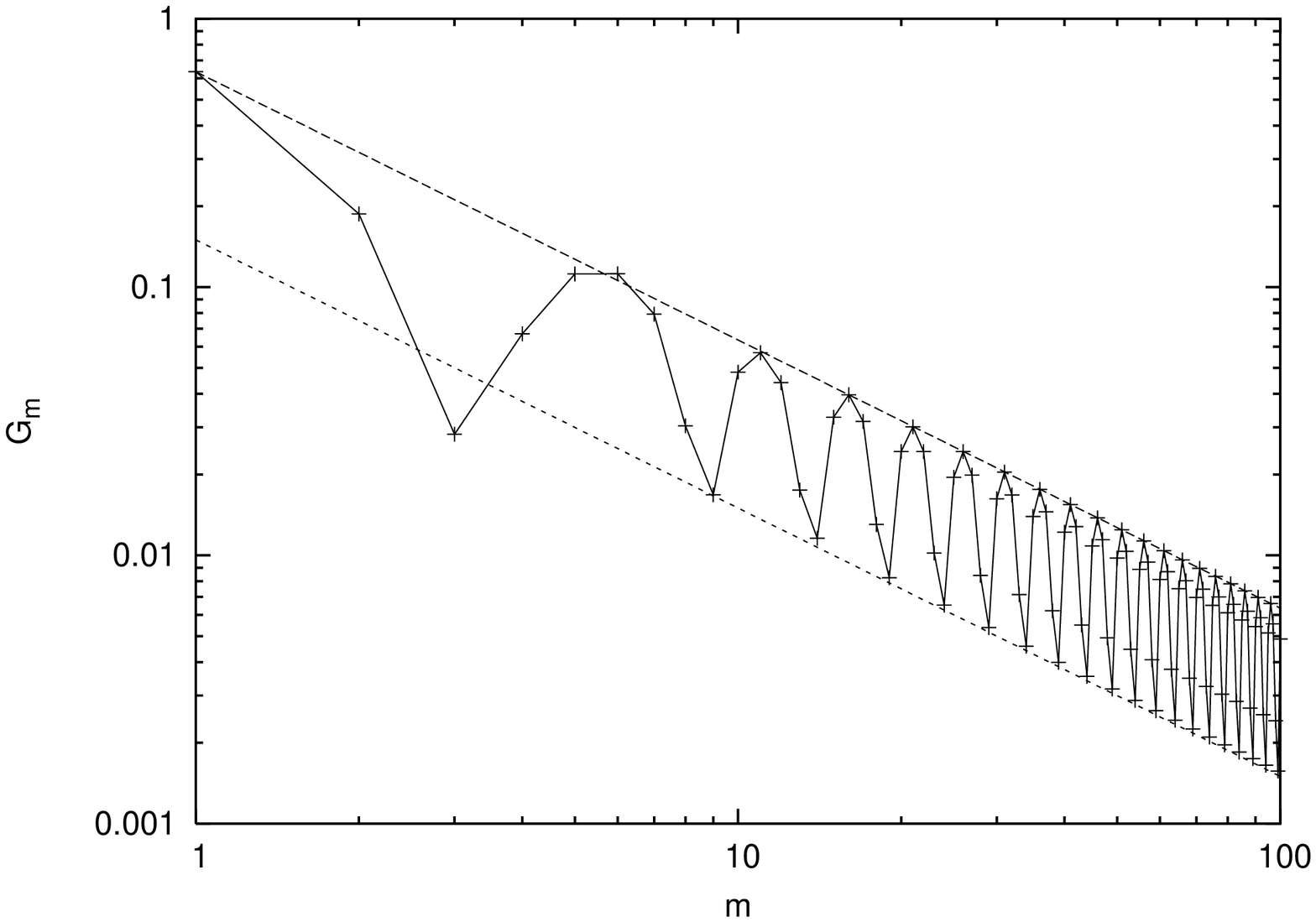}
\caption\protect{\label{fig:Gm}Plot of the elements of the Toeplitz
determinant for $\phi=0.8$ and $\lambda=1.3$. The oscillations themselves
depend on $\phi$ and $\lambda$ but the upper and the lower envelopes decay as 
$1/m$ for all $\lambda$ and $\phi$. The dotted lines are $0.635/m$ and 
$0.150/m$.}
\end{figure}

Let us now investigate whether the conditions required (other than zero 
index) for Szego's theorem to be valid are satisfied here or not. 
For $\phi=\pi$ (ground state, all modes unexcited), one can
make binomial expansion of the surd in $w$ (Eq. (11)) and conclude that for
any large positive integral value of $m$ (with $\lambda < 1$),
\be \mid \overline{G}_m \mid \; \propto \frac{\lambda^{m}}{\sqrt{m}} \ee
and
\be \mid \overline{G}_{-m} \mid \; \propto \frac{\lambda^{m}}{m\sqrt{m}} \ee
Since $\overline{G}_m$ and $\overline{G}_{-m}$ both vanish exponentially 
with $m$, the conditions,
\[ \sum_{m=-\infty}^{\infty}  \mid \overline{G}_m \mid < \infty \;\;\;\;\;\;\;\;
{\rm and} \;\;\;\;\;\;\;\;
\sum_{m=-\infty}^{\infty} \mid m \mid \mid \overline{G}_m \mid^2 < \infty \]
are valid. Since $\overline{c}(k) \neq 0$ for all values of $k$, the 
Hirshmann conditions \cite{MW3,FH1} for the validity of Szego's theorem is 
satisfied. (However, it is not clear whether the Devinatz 
condition \cite{MW3}, namely, the continuity of
\[ \sum_{m=1}^{\infty}  \left( \overline{G}_m\;e^{imk} - \overline{G}_{-m}\;e^{-imk} \right) \]
is true.)
For $\phi \ne \pi$, the asymptotic behaviour of $\mid \overline{G}_m \mid$
and $\mid \overline{G}_{-m} \mid$ depends on $\lambda$ and $\phi$ and is not
easy to track analytically. Numerical study shows that both these quantities 
are oscillatory (Fig.~\ref{fig:Gm}). 
The upper and lower envelopes of the oscillatory curve
decay as $1/m$ for all $\lambda$ and $\phi$ while the wavelength and 
amplitude of oscillation depends on the specific values of $\lambda$ and 
$\phi$. It is thus difficult to analyse the validity of Hirshmann or
Devinatz condition here. We do not proceed further with the question of
validity of the conditions for {\em applicability} of Szego's theorem and
rather choose to study the results obtained when we do apply the theorem
anyway. 

To apply Szego's theorem, the next step is to calculate the quantities 
$g_0$ and $g_{\pm m}$ defined as
\be g_0 = \frac{1}{2\pi} \int_0^{2\pi} dk \; \ln \overline{c}(k) \ee
\be  g_{\pm m} = \frac{1}{2\pi} \int_0^{2\pi} dk \; e^{\mp imk} \; \ln 
\overline{c}(k), \ee
where $m$ is any positive integer. These quantities can
be easily evaluated from Eq. (13), but we must be careful to choose the 
principal branch for $\ln \overline{c}(k)$, satisfying
\[ -i\pi < \ln \overline{c}(k) < i\pi. \]
This criterion is satisfied if we express the logarithm as,
\be \ln \overline{c}(k) = ik + \frac{1}{2} \ln \frac{1 + \lambda e^{-ik}}
{1 + \lambda e^{ik}} + i\pi\delta \tag{17a}\ee
with the phase factor
\[ \delta = \left\{ \begin{array}{rl}
                     0 & \mbox{for $0 < k < \phi$}\\
                     -1 & \mbox{for $\phi < k < 2\pi - \phi$}\\
                     -2 & \mbox{for $2\pi - \phi < k < 2\pi$}.\\
         \end{array}
 \right. \]

By expanding the logarithm on the right hand side of Eq. (17a) in a series 
and performing a term-by-term integration, we now get
\be g_0 = 0 \ee
and
\be g_{\pm m} = \pm \frac{(-\lambda)^m}{2m} \mp \frac{\cos(m\phi)}{m}. \ee

Next, we have to calculate the sum,
\be \sigma = \sum_{m=1}^{\infty} m g_m g_{-m} \ee
Using the standard result \cite{GR}
\be \sum_{n=1}^{\infty} \frac{1}{m}a^m \cos(m\alpha) = 
- \frac{1}{2} \ln \left(1 - 2 a \cos\alpha + a^2\right) \ee
(for $0 < \alpha < 2\pi$, $a^2 \le 1$) one obtains,
\[ \sigma = \frac{1}{4} \ln \left( 1 - \lambda^2 \right )  
+ \frac{1}{2}\ln \left(\frac{2\sin \phi}{1+2\lambda\cos\phi + \lambda^2} \right)
- \frac{1}{2} \sum_m \frac{1}{m}. \]
The last sum is divergent. Following the conjecture of Fisher and Hartwig
\cite{FH,bott,ehr,FH3}, 
we replace the upper limit of the sum by $\zeta n$ to obtain
\be \sum_{m=1}^{\zeta n} \frac{1}{m} = \gamma_E + \ln (\zeta n) \ee
where $\gamma_E$ is the Euler's constant (=0.5772). The constant $\zeta$ is 
independent of $\lambda$, $\phi$ and $n$. Its numerical value could not be
found analytically.

Szego's theorem now reads \cite{MW,FH,bott}
\[ {\overline{\det}}_n(\lambda) = \exp(n g_0 + \sigma), \] 
which, on substitution of the expressions of $g_0$ and $\sigma$ leads to
the final result 
\be C^z(n) = {\overline{\det}}_n(\lambda) = 
\frac{A\sqrt{\sin \phi}(1-\lambda^2)^{1/4}}
{\sqrt{1+2\lambda\cos\phi + \lambda^2}}\frac{1}{\sqrt{n}}, \ee
where $A=(1/\sqrt{\zeta/2})\exp(-\gamma_E/2)$. We have numerically evaluated 
this constant to be 0.590 (which corresponds to $\zeta=3.226$) by computing 
determinants upto a size of $200 \times 200$. 
The correlation function $C^z(n)$ decays algebraically with distance without 
any oscillation. 

\section{Spin Correlations for $\lambda > 1$}

We shall now evaluate the Toeplitz determinant of Eq. (7) for $\lambda > 1$.
Since the index of the generating function is not zero in this case (as
mentioned above), Szego's theorem cannot be applied directly. One has 
to follow instead the method of Wu \cite{MW2}. According to Eq. (8), the 
elements $G_m$ for some given value of $\lambda$ is related to the same for
$1/\lambda$ by,
\be G_m (\lambda) = G_{-m+1} (1/\lambda) \ee
Hence, the determinant for $\lambda=\lambda_0> 1$,
can be written as,
\be {\det}_n(\lambda_0) = \left | \begin {array}{ccccc}
  G_1 & G_0 & G_{-1} & \cdots & G_{-n+2} \\
  G_2 & G_1 & G_0 & \cdots & G_{-n+3} \\
  \cdots \cdots & & & & \\
  G_n & G_{n-1} & G_{n-2} & \cdots & G_1 \\
\end{array}      \right |_{\lambda=1/\lambda_0}
\ee
Let us consider the Wiener-Hopf system of equations,
\be \sum_{m=0}^n \; G_{k-m}(1/\lambda_0)\; x_m = \delta_{k,0} \ee
for $k=0$, 1, $\cdots n$. By Cramer's rule, 
\be x_n(1/\lambda_0) = (-1)^n \frac{{\det}_n(\lambda_0)}
{{\det}_{n+1}(1/\lambda_0)} \ee
This equation would give the value of the determinant for $\lambda > 1$ in 
terms of the determinant for $\lambda < 1$ provided we know $x_n$ for 
$\lambda < 1$. 

To proceed further, we switch over to the convention that the root which 
ensures $ c(\pi) > 0$ will be chosen and solve the Wiener-Hopf equations
\be \sum_{m=0}^n \; \overline{G}_{k-m}(1/\lambda_0)\;
\overline{x}_m = \delta_{k,0}.\ee
The solution for $\overline{x}_n$ as found by Wu \cite{MW4} is  
\be  \overline{x}_n = \frac{1}{2\pi} \int_0^{2\pi} dk \; e^{ink} \; l(k), \ee
with $l(k)$ defined through \cite{FH2}
\be \ln l(k) = \sum_{m=1}^{\infty} g_{-m}e^{imk} - 
\sum_{m=0}^{\infty} g_{m}e^{-imk}. \ee
Putting here the expressions of $g_0$ and $g_{\pm m}$ (Eqs. (18), (19))
directly and using (21), we get
\begin{eqnarray}
\overline{x}_n & = & \frac{1}{2\pi} \int_0^{2\pi} dk \; e^{ink}\;
\sqrt{1+\lambda^2 + 2\lambda\cos k} \; \times \nonumber \\ 
& & \exp\left [ \sum_{m=1}^{\infty} \frac{2}{m}
\cos(m\phi) \cos(mk) \right]\end{eqnarray}
We rewrite this as
\begin{eqnarray*}
&\overline{x}_n & = \frac{1}{\pi} \int_0^{\pi} dk \; \cos(nk)\;
\sqrt{1+\lambda^2 + 2\lambda\cos k} \; \times \\
& & \exp\left [ \sum_{m=1}^{\infty} \frac{1}{m}\{\cos(m\phi+mk) 
+ \cos(m\phi-mk) \} \right]\end{eqnarray*}
and note that the formula (21) may be applied to $\sum (1/m)\cos(m\phi+mk)$ 
but not to $\sum (1/m)\cos(m\phi-mk)$ since the former is convergent in the 
range of integration while the latter diverges at $k = \phi$. Thus we have,
\be \overline{x}_n = \frac{1}{\pi} \int_0^{\pi}
\; dk \; \frac {\cos(nk)\sqrt{1+\lambda^2 + 2\lambda\cos k}}
{2\sin\frac{\phi+k}{2}} \;e^{{\mathcal S}} \ee
where,
\be {\mathcal S} =  \sum_{m=1}^{\infty} \frac{1}{m} \cos(k-\phi)m \ee
For large $n$, the term $\cos(nk)$ oscillates rapidly and the non-trivial
contribution to the integral comes only from the points where the rest of the
integrand diverges. Hence, it suffices to integrate only over a narrow
region (of width $2\epsilon \ll 1/n$) around the point $k=\phi$ and obtain
\be \overline{x}_n = \frac{1}{2\pi}
\frac{\sqrt{1+\lambda^2 + 2\lambda\cos (\phi)}}{\sin(\phi)}
\int_{-\epsilon}^{\epsilon} \cos(nx+n\phi) \; e^{{\mathcal S}} \; dx \ee
where we have taken the smoothly varying part out of the integration and put
$x=k-\phi$.
Following the Fisher-Hartwig conjecture,
we now replace (``heuristically'') the upper limits in the sum ${\mathcal S}$
(see Eq. (33)) by $\zeta^{\prime}n$ and note that for 
$\zeta^{\prime} \sim 1$, we must have
$xm \ll 1$ for all $m$ in the range of integration. The sum ${\mathcal S}$ may
then be approximated as 
\be {\mathcal S} =  \sum_{m=1}^{\zeta^{\prime}n} \left [\frac{1}{m} -
\frac{1}{2} x^2 m \right]. \ee
Substituting Eq. (22) in the first summation, we obtain,
\be e^{\mathcal S} = 2 \sqrt{\pi} e^{\gamma_E} \;
\left[ \frac{\zeta^{\prime} n}{2\sqrt{\pi}} e^{-x^2 (\zeta^{\prime}n/2)^2} 
\right] \ee
As $n \rightarrow \infty$, the portion within $[ \cdots ]$ becomes
the Dirac delta functon $\delta(x)$ and the expression (34) for
$\overline{x}_n$ becomes
\be \overline{x}_n = \frac{e^{\gamma_E}}{\sqrt{\pi}} \; 
\frac{\sqrt{1+\lambda^2 + 2\lambda\cos (\phi)}}{\sin(\phi)} \; \cos(n\phi) \ee
The unknown constant $\zeta^{\prime}$ (happily) does not appear in this final 
expression. 
Since $\overline{x}_n = x_n$, Eqs. (12a), (23), and (27) now give
our final expression for the correlation function for $\lambda > 1$ :
\begin{eqnarray}
C^z(n) & = & {\det}_n(\lambda) = (-1)^n \;x_n(1/\lambda) \;
{\det}_{n+1}(1/\lambda) \nonumber \\
 & = & \frac{A e^{\gamma_E}}{\sqrt{\pi}\sin(\phi)}
\left( 1-\frac{1}{\lambda^2} \right)^{1/4} \frac{\cos(\pi+\phi)n}{\sqrt{n}}.
 \end{eqnarray}
(The constant $A$ is defined below Eq. (23).)
This correlation function decays algebraically with distance and also
oscillates with distance with wave-vector $(\pi + \phi)$.

Before we conclude this section, we point out that although Eq.(29) for 
$\overline{x}_n$ as derived by Wu \cite{MW4} is applicable here, 
there is a major 
difference between our case and the one studied by Wu. In our case, 
$x_n$ does not vanish in the limit $n \rightarrow \infty$ and all we want to
evaluate is the zero-th order term in $n$. But in Wu's treatment, 
$\lim_{n \rightarrow \infty} x_n$ is zero and the basic job was to evaluate 
the leading terms for large $n$. 

\section{Discussions}

(i) If we look at the spin-spin correlation in the ground state of transverse
Ising model, then we find that it is only for $\lambda=1$ that the correlation
decays algebraically with distance. But for the type of excited states we study
(see Eq. (5)), the algebraic behaviour is seen over a larger range, namely, for 
$\lambda < 1$ as well as $\lambda > 1$ and any $\phi$ in the range 
$0 < \phi < \pi$. \\
(ii) For handling the divergent sums that arise in the calculation of 
correlation function, Fisher-Hartwig prescription is very much useful for the
excited states as it was for the ground state. Actually, the index for 
algebraic decay can be derived from the generating function itself, without any
further calculation. Fisher and Hartwig have shown that
for generating function having multiple singularities, if $\arg[c(k)]$ jumps
by an amount $2\pi i \alpha_j$ at the $j$-th singularity, then the determinant
will decay algebraically with distance with an exponent $\sum_j \alpha_j^2$.
In our case, the generating function of Eq. (10) has two jumps, at each of
which $\alpha_j$ is $1/2$, and the index is $1/2$, obeying this conclusion.
Moreover, if we consider an excited state where (in contrast to Eq. (5)) the
excited modes extend from $k=\phi_1$ to $\phi_2$, then (for $\phi_2 \ne \pi$)
there will be four singularities in the generating function, at each of which
$\alpha_j$ is $1/2$. The correlation will then decay as $1/n$. \\

\begin{acknowledgements}
The work of one author (AKC) was supported by UGC fellowship. 
\end{acknowledgements}


\begin{references}

\bibitem{bkc_book} B.K.~Chakrabarti, A.~Dutta and P.~Sen, {\em Quantum Ising 
Phases and Transitions in Transverse Ising Models} (Springer-Verlag, Berlin, 
Heidelberg) 1996. 

\bibitem{rieger} G.~Uimin and H.~Rieger, Z. Phys. B {\bf 101}, 597 (1996). 

\bibitem{amit}  A.~Dutta and D.~Sen, Phys. Rev. B {\bf 67}, 094435 (2003). 

\bibitem{cd1} A. K. Chandra and S. Dasgupta, arXiv:cond-mat/0612144, 
Phys. Rev. E {\bf 75}, 021105 (2007).

\bibitem{LSM} E. Lieb, T. Schultz and D. C. Mattis, Annals of Phys. {\bf 16},
 407 (1961).
 
\bibitem{pf} P. Pfeuty, Annals of Phys. {\bf 57}, 79 (1970).

\bibitem{mattis} D. C. Mattis, {\em The Theory of Magnetism}, Vol. II
(Springer-Verlag, Berlin, Heidelberg) 1985, Sec. 3.6.

\bibitem{MC} B.M. McCoy, Phys. Rev. {\bf 173}, 531 (1968).                      

\bibitem{LSM1} See Ref \cite{LSM} Eq. (2.88).

\bibitem{MW} For an excellent elaborate discussion (for a physicist) on
Toeplitz determinant and Szego's Theorem, see B.M.~McCoy and T.T.~Wu, {\em The
Two-Dimensional Ising Model} (Harvard University Press,
Cambridge, Massachusetts) 1973 Chapter X.

\bibitem{FH} M.E.~ Fisher and R.E.~Hartwig, Adv. Chem. Phys. {\bf 15}, 333
(1968)
 
\bibitem{bott} A.~Bottcher, J. Stat. Phys. {\bf 78}, 575 (1995).
 
\bibitem{ehr} T. Ehrhardt, Operator Th : Advances and App. {\bf 124} 217
(2001).

\bibitem{MW1} McCoy and Wu, \cite{MW}, pages 186, 244.

\bibitem{MW5} McCoy and Wu \cite{MW}, p. 250; T.T.~Wu, Phys. Rev.
{\bf 149}, 380 (1966), see p.381.

\bibitem{MW3} McCoy and Wu \cite{MW}, p. 225.
 
\bibitem{FH1} Fisher and Hartwig \cite{FH}, p. 338, Theorem 2.

\bibitem{GR} I.S.~Gradshteyn and I.M.~Ryzhik, {\em Table of Integrals, Series,
and Products} (Academic Press, New York) 1980 Entry No. (1.448 2).

\bibitem{FH3} Fisher and Hartwig's `conjecture' is however, no longer a
conjecture, since it has been proven by E.L.~Basor, Indiana Math. J. {\bf 28},
975 (1979).

\bibitem{MW2} McCoy and Wu, \cite{MW}, Chapter XI; also in T.T.~Wu, Phys. Rev.
{\bf 149}, 380 (1966) .

\bibitem{MW4} McCoy and Wu \cite{MW}, p. 253, Eq. (2.27); Wu \cite{MW2}
Eq. (2.27).

\bibitem{FH2} Fisher and Hartwig \cite{FH}, p. 341, Eq. (42). The $m=0$ term
must be added to the second summation there. 

\end{references}
\end{document}